# Probing polarization response of monolayer cell cultures with photon entanglement

L. Zhang[1,2,3]*, V. R. Besaga[2], P. Rühl[4], C. Zou[1,3], S. H. Heinemann[4,5], Y. Wang[1,3], and F. Setzpfandt[2,6]

__________

[1] Institute of Microelectronics of the Chinese Academy of Sciences, 100029 Beijing, China

[2] Institute of Applied Physics, Abbe Center of Photonics, Friedrich Schiller University Jena, 07745 Jena, Germany

[3] University of the Chinese Academy of Sciences, 101408 Beijing, China

[4] Center for Molecular Biomedicine, Department of Biophysics, Friedrich Schiller University Jena, 07745 Jena, Germany

[5] Jena University Hospital, 07745 Jena, Germany

[6] Fraunhofer Institute for Applied Optics and Precision Engineering IOF, 07745 Jena, Germany

*Correspondence

Luosha Zhang

[1] Institute of Microelectronics of the Chinese Academy of Sciences, 100029 Beijing, China

[2] Institute of Applied Physics, Abbe Center of Photonics, Friedrich Schiller University Jena, 07745 Jena, Germany

[3] University of the Chinese Academy of Sciences, 101408 Beijing, China

Email: luosha.zhang@uni-jena.de

*((Optional footnotes:*

*L. Zhang and V. R. Besaga have contributed equally to this work.*

*L. Zhang and V. R. Besaga should be considered joint first author))*

**Abstract**

This study addresses the critical need for high signal-to-noise ratio in optical detection methods for biological sample discrimination under low-photon-flux conditions to ensure accuracy without compromising sample integrity. We explore polarization-based probing, which often excels over intensity modulation when assessing a specimen's morphology. Leveraging non-classical light sources, our approach capitalizes on sub-Poissonian photon statistics and quantum correlation-based measurements. We present a novel, highly sensitive method for probing single-layer cell cultures using entangled photon pairs. Our approach demonstrates capability in monolayer cell analysis, distinguishing between two types of monolayer cells and their host medium. The experimental results highlight our method's sensitivity, showcasing its potential for biological sample detection using quantum techniques, and paving the way for advanced diagnostic methodologies.

# 1    INTRODUCTION

Compared to sensing with classical light, non-classical states of light can enable lower fluctuations in the photon number [1], enhanced signal-to-noise ratio (SNR) [2] and thus are suitable for low-light level sensing and imaging [3]. In particular, non-classical light can have sub-Poissonian statistics, where the fluctuation in the photon number is below the classical shot-noise limit [4]. This reduced fluctuation is particularly advantageous in applications requiring high-precision sensing or measurements, where it leads to improved SNR and thus higher sensitivity.

In addition to fundamentally surpassing classical limits for the SNR using lower fluctuations of the photon number in non-classical light, typically used measurement techniques for non-classical states of light, such as correlation measurement, can also lead to an effective reduction of environmental noise caused by detectors or scattered light in the system. For instance, Kalashnikov, D.A., *et al.* [5] compared the SNR in spectroscopic measurements using non-classical and classical light sources. With their experimental setting, the SNR using a non-classical light source was 70,000 times greater than that of spectroscopy with classical light.

In low-light level sensing and imaging, thermal and pseudo-thermal light sources are not optimal for measurement due to their relatively high intensity fluctuations. For non-classical light sources, the inherent uncertainty is given by the Heisenberg limit, which reduces the system's measurement uncertainty by a factor of $\sqrt{N}$ compared to the best classical case, where N is photon number involved in interaction with the object under study [1].

Due to the features described above, non-classical states of light promise to have significant advantages in cell-culture probing because of benefits from enhanced SNR, noise reduction in weak-light measurement that will not damage live cells [6].

Polarimetry, the measurement and analysis of the polarization of light, can also be highly effective in biological sample probing due to its non-invasive and sensitive nature. It was implemented for diverse application such as microstructural feature characterization [7], diagnostic early gastric cancer detection [8], studying cell and tissue properties during zebrafish embryonic development [9], evaluating the freshness of ingredients [10], as well as surgical and further diagnostic applications [11]. In all the mentioned studies, classical light sources were used.

In this work, we endeavor to increase sensitivity and precision with enhanced SNR in polarimetric sensing by employing non-classical states of light. We studied the feasibility of applying polarization-entangled photon pairs for probing monolayer cultures of mammalian cells with weak polarization response. We experimentally demonstrate the potential of such an approach by distinguishing different cell cultures based on the quantum state changes. Our results promise a practical potential in biological sample characterization.

## 2    EXPERIMENTAL REALIZATION

### 2.1    Cell Culture Preparation

#### 2.1.1    Cell Culture

In this study the following mammalian cell lines were used: CHO-K1 Chinese hamster ovary cells (CHO) were obtained from the German Collection of Microorganisms and Cell Cultures (Braunschweig, Germany) and HEK293T human embryonic kidney cells (HEK) were acquired from the Centre for Applied Microbiology and Research (Porton Down, Salisbury, UK). Cells were cultivated in DMEM/F12 (Thermo Fisher Scientific, Waltham, Massachusetts, USA) and maintained in a humidified incubator with 5% $CO_2$ at 37°C. Representative images of cultivated confluent cell cultures (Zeiss, with EC Epiplan-Neofluar 20×/0.5 HD DIC M27 objective) are shown in Figure 1.

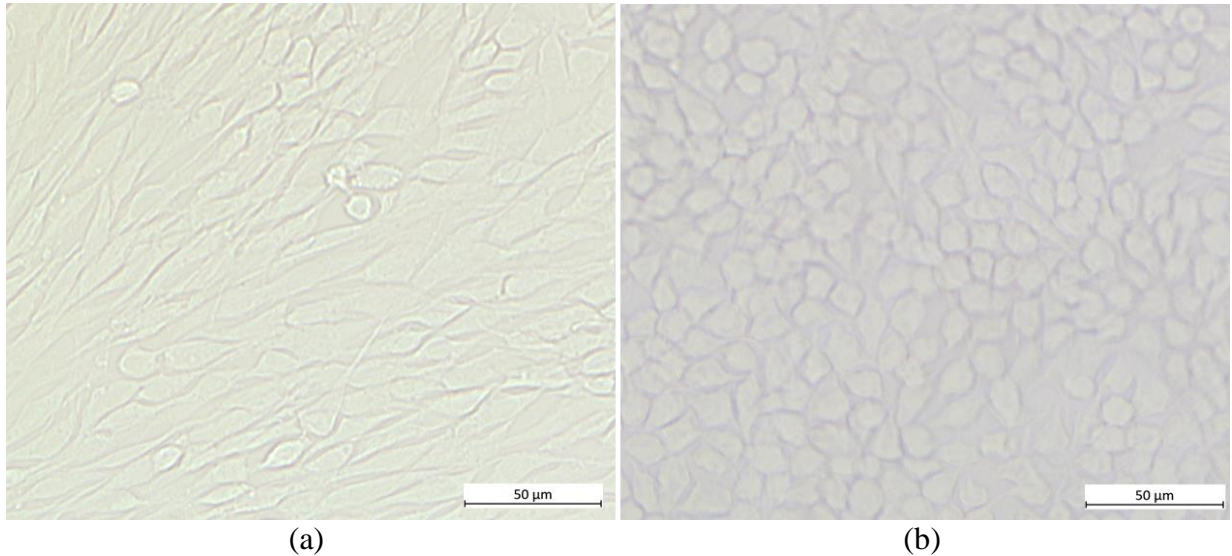

(a)                  (b)

Figure 1. Representative images of cell cultures used in this study. (a) Chinese hamster ovary cells (CHO-K1), (b) human embryonic kidney cells (HEK293T).

2.1.2    Preparation of Fixed Cells

Cells were grown on 12-mm borosilicate glass coverslips (Epredia, Breda, Netherlands) in a 24-well plate (Greiner Bio-One, Frickenhausen, Germany) until they formed a confluent monolayer. For fixation, the cell culture medium was removed. The cells were washed three times with 1 ml of phosphate-buffered saline (PBS, pH 7.4), fixed for 5 minutes with a 4% paraformaldehyde solution (PFA, pH 6.9) at ambient temperature, and kept in PBS. The fixed samples were rinsed once with deionized $H_2O$, and the coverslips were mounted on $76 \times 26$ $mm^2$ glass slides (Thermo Scientific) using Immu-Mount (Thermo Scientific). For reference measurements, samples containing only Immu-Mount, were prepared as reference samples, where glass coverslips were washed once with deionized $H_2O$ and mounted with Immu-Mount (host material of cells) on a glass slide, to investigate the relative polarization response caused by the CHO and HEK cell monolayers. For each condition, multiple areas from 2 coverslips were recorded.

2.2    Experimental Realization

To study the feasibility of the polarimetric probing of monolayer of cell cultures with polarization-entangled photon pairs we utilized the experimental setup sketched in Figure 2.

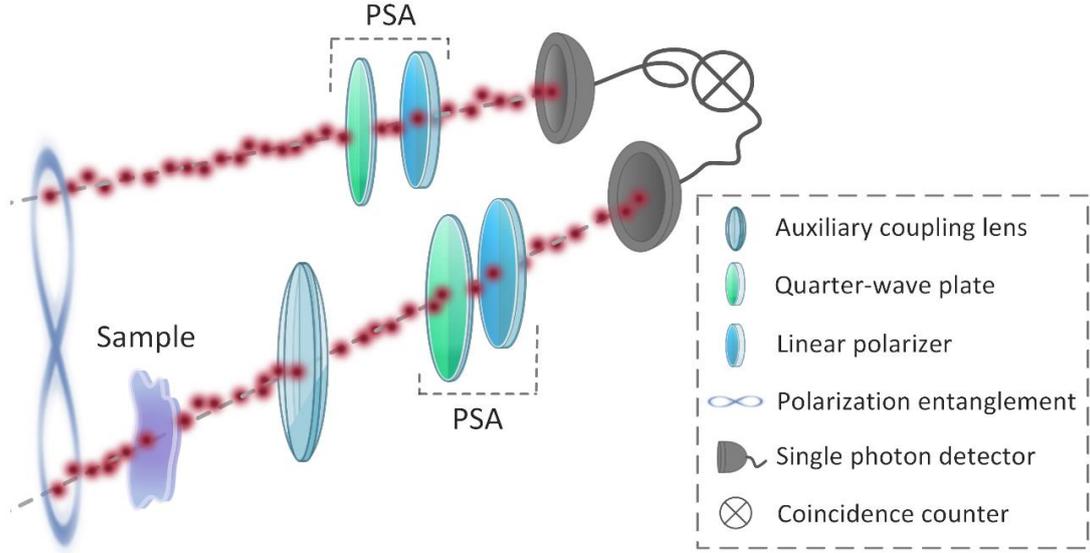

Figure 2. Conceptual sketch of the optical setup used in the experiments. A pair of polarization-entangled photons is split into two channels. Polarization state analyzers (PSA) that consist of a quarter wave plate and a linear polarizer are introduced in each channel before the detection for polarization projective measurements. A sample under study is introduced into one of the channels (signal channel). A low-focusing auxiliary coupling lens is introduced after the sample to enhance collection of the photons scattered by the sample towards the detector. The second path acts as a reference channel. Coincidence events between the channels are measured with a time tagging device.

The wavelength-degenerate polarization-entangled photon pairs at 810 nm are generated with two periodically poled potassium titanyl phosphate (KTiOPO$_4$) crystals with type-II phase-matching within a polarization Mach-Zehnder interferometer [12] (not depicted in the figure). The quantum state of the entangled source used in the study ideally is $|\Phi\rangle = 1/\sqrt{2}\,(|HV\rangle + |VH\rangle)$, where H and V denote horizontal and vertical polarization, respectively.

To characterize the state emitted by our source and also the changes induced by the cell samples under test, we employed quantum-state tomography in the polarization degree of freedom. Here, a polarization state analyzer (PSA) that consists of a quarter-wave plate (QWP) and a linear polarizer (LP) was used in each channel of the instrument to project the photons into different states. Coincidence events between the two channels were measured using fiber-coupled single-photon detectors and a time-tagging device.

To obtain the complete information about the quantum state, correlations were measured for 16 distinct combinations of polarization projections [13], which are described in Table 1 by the angles of the QWP's fast axis and the transmission axis of the LP with respect to the global vertical orientation in the laboratory. $|R\rangle$, $|L\rangle$, $|D\rangle$ correspond to right circular, left circular and diagonal polarization, respectively. $|HV\rangle$ corresponds to the case when a photon with horizontal polarization is projected in the reference channel, and vertical in the signal channel. The rest of the combinations follow similarly.

Table 1. The 16 distinct combinations of polarization projections used for quantum polarization state tomography. 0° corresponds to global vertical orientation in the laboratory.

| No. | Reference channel | Signal channel | QWP$_R$ | LP$_R$ | QWP$_S$ | LP$_S$ |
|---|---|---|---|---|---|---|
| 1 | $|H\rangle$ | $|H\rangle$ | 45° | 0° | 45° | 0° |
| 2 | $|H\rangle$ | $|V\rangle$ | 45° | 0° | 0° | 0° |
| 3 | $|V\rangle$ | $|V\rangle$ | 0° | 0° | 0° | 0° |
| 4 | $|V\rangle$ | $|H\rangle$ | 0° | 0° | 45° | 0° |
| 5 | $|R\rangle$ | $|H\rangle$ | 22.5° | 0° | 45° | 0° |
| 6 | $|R\rangle$ | $|V\rangle$ | 22.5° | 0° | 0° | 0° |
| 7 | $|D\rangle$ | $|V\rangle$ | 22.5° | 45° | 0° | 0° |
| 8 | $|D\rangle$ | $|H\rangle$ | 22.5° | 45° | 45° | 0° |
| 9 | $|D\rangle$ | $|R\rangle$ | 22.5° | 45° | 22.5° | 0° |
| 10 | $|D\rangle$ | $|D\rangle$ | 22.5° | 45° | 22.5° | 45° |
| 11 | $|R\rangle$ | $|D\rangle$ | 22.5° | 0° | 22.5° | 45° |
| 12 | $|H\rangle$ | $|D\rangle$ | 45° | 0° | 22.5° | 45° |
| 13 | $|V\rangle$ | $|D\rangle$ | 0° | 0° | 22.5° | 45° |
| 14 | $|V\rangle$ | $|L\rangle$ | 0° | 0° | 22.5° | 90° |
| 15 | $|H\rangle$ | $|L\rangle$ | 45° | 0° | 22.5° | 90° |
| 16 | $|R\rangle$ | $|L\rangle$ | 22.5° | 0° | 22.5° | 90° |

First, we characterized the quantum state emitted by our source using quantum tomography. The real and imaginary parts of the density matrix calculated from the measured correlations are depicted in Figure 3. The measured density matrix shows a good agreement with the one expected for the state $|\Phi\rangle$ (marked with dash lines), where the two diagonal elements $|HV\rangle\langle HV|$ and $|VH\rangle\langle VH|$, describing the probabilities of measuring the respective combination of photons, have magnitudes close to 0.5. Their mutual coherences in the corresponding off-diagonal elements also are close to 0.5. Overall, the fidelity of the measured with the ideal quantum state is around 0.95, the slight deviations stems from an asymmetry between the two paths within the source.

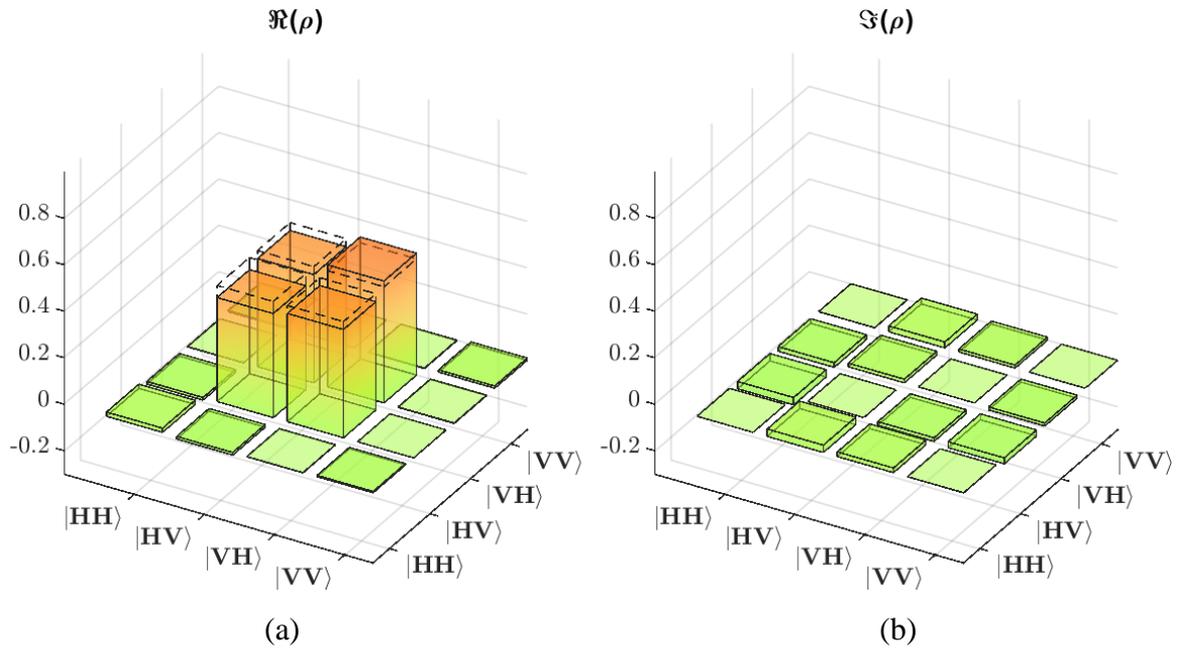

Figure 3. (a) Real and (b) imaginary part of the representative measured density matrix characterizing the quantum state used in the experiments.

The generated photon pairs were divided into two paths so that one of the photons interacts with the sample, while its partner acts as a reference; we denote the first path as signal channel and the other as reference channel.

In the experiments the integral polarization response of the sample is probed, and the sample is illuminated with a collimated beam of approximately 1 mm diameter. The total number of photons incident on the sample is around 500,000 photons per second which, considering the illumination spot area on the sample, corresponds to photon flux of about $6.4\times10^{11}$-photons/m$^2$/s.

No focusing or imaging optics is used before the sample. The integral measurement without spatial resolution governs also the choice of the samples. To examine the sensitivity of the method with respect to sample morphology, different shapes of cell types were selected, and the cell cultures have been prepared as monolayers.

Considering the weak interaction of the specimen under study with light, the changes in polarization induced by the sample can be smaller than signal loss due to back-reflection or absorption in glass substrates, scattering, etc. To compensate for the potential losses caused by scattering, we introduced a low-focusing auxiliary coupling lens after the sample in the signal channel.

To detect the changes in the polarization state of the transmitted photons induced by the cell samples, the quantum state was measured before and after the introduction of samples by obtaining the number of correlations events within 10-s integration time and 3-ns coincidence window for the 16 distinct polarizer projection combinations shown in Table 1. The obtained coincidence counts were combined into vectors $\psi_{sample}$ for the measurement with the sample and $\psi_{reference}$ for the measurement without the sample as:

$$\psi_{sample} = \begin{bmatrix} c_{HH}^s \\ c_{HV}^s \\ ... \\ c_{RL}^s \end{bmatrix}, \quad \psi_{reference} = \begin{bmatrix} c_{HH}^r \\ c_{HV}^r \\ ... \\ c_{RL}^r \end{bmatrix}. \quad (2)$$

In total, 18 samples of CHO, 14 samples of both HEK and Immu-Mount were measured across both two prepared cell culture areas.

2.3  Data Analysis

The changes in the coincidence counts (here referred to as "coincidence"), which are caused by the presence of the samples, across the 16 distinct polarizer projection combinations was calculated as:

$$\psi_{variation} = \psi_{sample} - \psi_{reference}. \quad (3)$$

The averaged coincidence and standard error of the mean (SEM) for the measurements were calculated.

The data distribution of the measurements was visualized in box plots. The box area illustrates that 50% of the data points fall within the marked range, the bottom and the top of the box represent the 25$^{th}$ and 75$^{th}$ percentile of the data, respectively.

Wilcoxon rank-sum test, also known as the Mann-Whitney U test [14], was employed to evaluate whether the measured cell types can be distinguished from each other, where the test is used to determine if there are statistically significant differences between two independent groups. As a measure, we used the p value, where a value below 0.05 indicates that the two compared groups are distinguishable [15].

A 4×4 matrix representing the impact of the sample under study, further referred to as a change matrix M with elements $m_{ij}$, was calculated for all cell cultures and the host medium using method employed to calculate the density matrix in quantum polarization state tomography [13]. For this, the values of the vector $\psi_{variation}$ were used as inputs for tomographic reconstruction algorithm including maximum likelihood estimation [15]. The matrix depicts the alteration in polarization of the photons transmitted by the sample.

## 3    RESULTS AND DISCUSSION

When a sample containing either the cell culture grown as a monolayer or a host material only is introduced into the optical path of photons in the signal channel, different phenomena of light-matter interaction take place concurrently. In order to address the effects of our cell samples on the polarization of the incident photons, the cell sample response is analyzed relative to the reference sample, i.e., the host medium Immu-Mount. If the suggested approach of probing with polarization-entangled photon pairs is sensitive enough, the response from the monolayer cell samples should be distinct from the response for host medium alone. Furthermore, if this response is related to specific properties of the different cells, e.g., the sample morphology, the polarization response should also be different for the two investigated cell types, whose cells have different shapes and the monolayers having different textures. In the following, we address these questions with statistical tests on the measured data.

3.1 Averaged Coincidence and SEM of the 16 Polarization Projections

First, we analyze the outcome from simple polarization projective measurements. To this end, the averaged coincidences and the SEM for all measurements were calculated for each of the 16 distinct combinations of polarization projections shown in Table 1.

The results are displayed in Figure 4, where the labels on the X axis correspond to the polarization combinations as described in Table 1. In the figure, the measurements for all 18 CHO,14 HEK, and 14 Immu-Mount areas under study contributed to the averaged coincidences changes caused from sample present and SEM. The solid markers represent averaged coincidence and same color shade areas represent corresponding values of SEM.

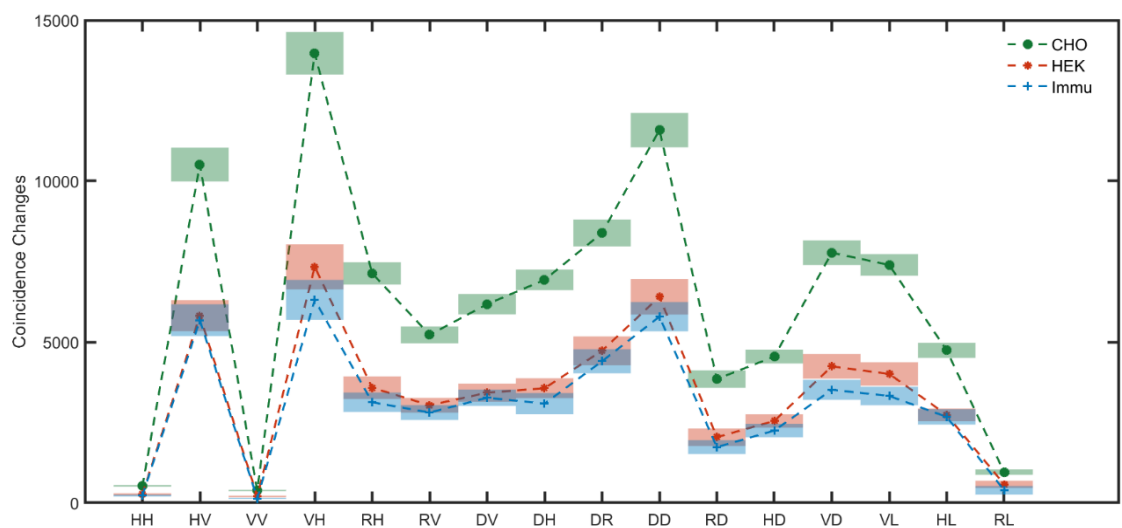

Figure 4. The averaged change in the coincidence counts and standard error of the mean for polarization projective measurements.

The values of averaged coincidence and SEM are used in the following for testing whether the sample presence can be detected and whether different cell types can be differentiated from each other.

3.2 Discrimination of Prepared Cell Types and Host Medium

Wilcoxon rank-sum test was performed for all 18 samples of CHO, 14 samples of HEK, and 14 samples of Immu-Mount using values of coincidences $\psi_{variation}$. The p values from Wilcoxon rank-sum test are displayed in Table 2, in which p values between CHO and Immu-Mount, between CHO and HEK, and between HEK and Immu-Mount are contained in the second, third, and fourth columns, respectively.

From the second column, one can clearly see that all 16 distinct combinations of polarization projections differ significantly ($p < 0.05$ for all) between CHO cells and the mounting medium, which means there are significant differences between these two sample types. This indicates the potential of distinguishing CHO cells from the homogeneous host medium using polarization measurements.

To study whether the detected polarization response is explained by solely scattering of photons on the cells or indeed by the polarization selectivity in the transmission due to the morphology of the sample, we investigate whether presence of another cell type (HEK) can be also detected and whether the method shows sensitivity to different cell types. Therefore, the test was performed between monolayer CHO and HEK cell cultures, and the third column of Table 2 shows that the p values are all far below 0.05, which indicates the potential to distinguish these two cell types.

At the same time, this does not hold true for the testing results between HEK and Immu-Mount samples, seen in fourth column, which indicates that only employing coincidences of the 16 distinct combinations of polarization projections cannot distinguish HEK cells from the host medium and further methods need to be employed. Therefore, we performed tomographic reconstructions for all the three cell types with the 16 distinct combinations of polarization projections to examine whether the difference between HEK and Immu-Mount can be extracted from elements of the change matrix M.

Table 2. Wilcoxon rank-sum test p values for 16 combinations of polarization projective measurements.

| Projections | CHO & Immu-Mount | CHO & HEK | HEK & Immu-Mount |
|---|---|---|---|
| HH | $6.17 \times 10^{-6}$ | $4.11 \times 10^{-5}$ | $6.08 \times 10^{-1}$ |
| HV | $7.01 \times 10^{-5}$ | $4.92 \times 10^{-5}$ | 1 |
| VV | $9.13 \times 10^{-6}$ | $5.86 \times 10^{-5}$ | $1.11 \times 10^{-1}$ |
| VH | $9.15 \times 10^{-6}$ | $2.85 \times 10^{-5}$ | $3.30 \times 10^{-1}$ |
| RH | $7.53 \times 10^{-6}$ | $2.85 \times 10^{-5}$ | $3.83 \times 10^{-1}$ |
| RV | $7.01 \times 10^{-5}$ | $1.95 \times 10^{-4}$ | $5.38 \times 10^{-1}$ |
| DV | $5.88 \times 10^{-5}$ | $1.40 \times 10^{-4}$ | $5.38 \times 10^{-1}$ |
| DH | $9.15 \times 10^{-6}$ | $1.96 \times 10^{-5}$ | $3.56 \times 10^{-1}$ |
| DR | $3.42 \times 10^{-5}$ | $9.93 \times 10^{-5}$ | $8.37 \times 10^{-1}$ |
| DD | $1.34 \times 10^{-5}$ | $3.42 \times 10^{-5}$ | $3.30 \times 10^{-1}$ |
| RD | $3.42 \times 10^{-5}$ | $1.95 \times 10^{-4}$ | $5.05 \times 10^{-1}$ |
| HD | $1.35 \times 10^{-5}$ | $3.42 \times 10^{-5}$ | $3.05 \times 10^{-1}$ |
| VD | $6.18 \times 10^{-6}$ | $4.92 \times 10^{-5}$ | $2.00 \times 10^{-1}$ |
| VL | $5.07 \times 10^{-6}$ | $3.42 \times 10^{-5}$ | $2.00 \times 10^{-1}$ |
| HL | $9.93 \times 10^{-5}$ | $8.35 \times 10^{-5}$ | $9.18 \times 10^{-1}$ |
| RL | $3.75 \times 10^{-4}$ | $1.20 \times 10^{-2}$ | $3.56 \times 10^{-1}$ |

3.2 Change Matrixes of All Samples

The averaged change matrixes of all samples of CHO, HEK and Immu-Mount are shown in Figure 5. The SEM of each averaged diagonal element is negligible and therefore is not shown in the figure. Dashed lines in the matrixes display the value 0.5 as a reference to compare. The left column displays the real part of the change matrix while the right column displays its imaginary part.

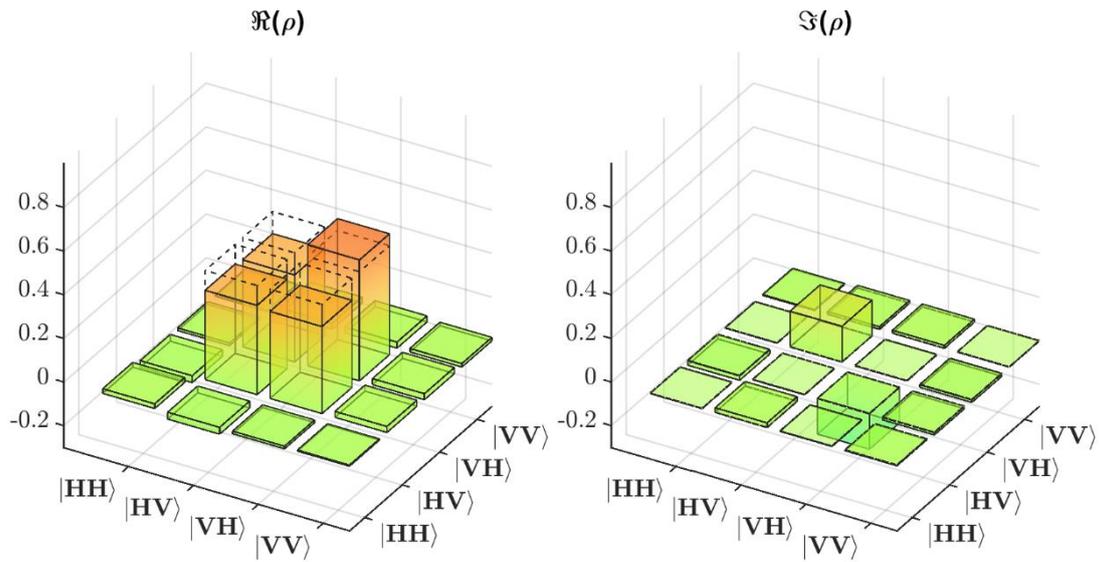

(a) CHO

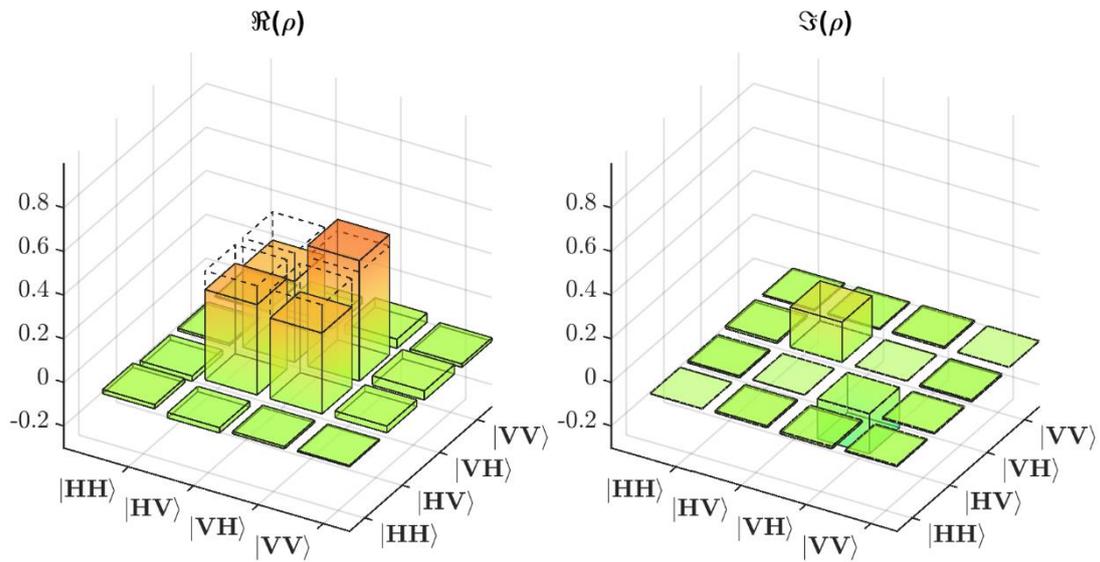

(b) HEK

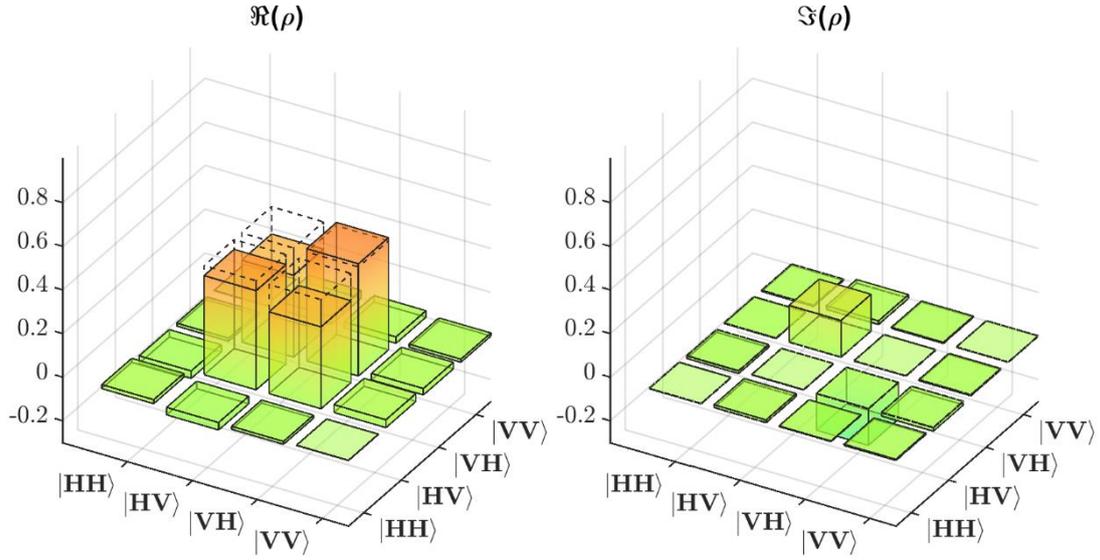

(c) Immu-Mount

Figure 5. Change matrixes of (a) CHO and (b) HEK monolayer cell cultures, and (c) host medium (Immu-Mount) only.

In order to compare three type cell cultures together, related elements were picked out for further analysis. Because of the polarization quantum state of the utilized entangled source, most counts and also maximal changes are expected for the four elements: $m_{2,2}$, $m_{2,3}$, $m_{3,2}$, $m_{3,3}$. Hence, only these four elements were considered for subsequent analysis.

Box plots of the four elements are shown in Figure 6. Due to the normalization during tomographic reconstruction, $m_{2,3}$, $m_{3,2}$ are complex conjugates of each other. It is therefore sufficient to analyze only $m_{2,3}$, which is analyzed and plotted separately with its real part and imaginary part.

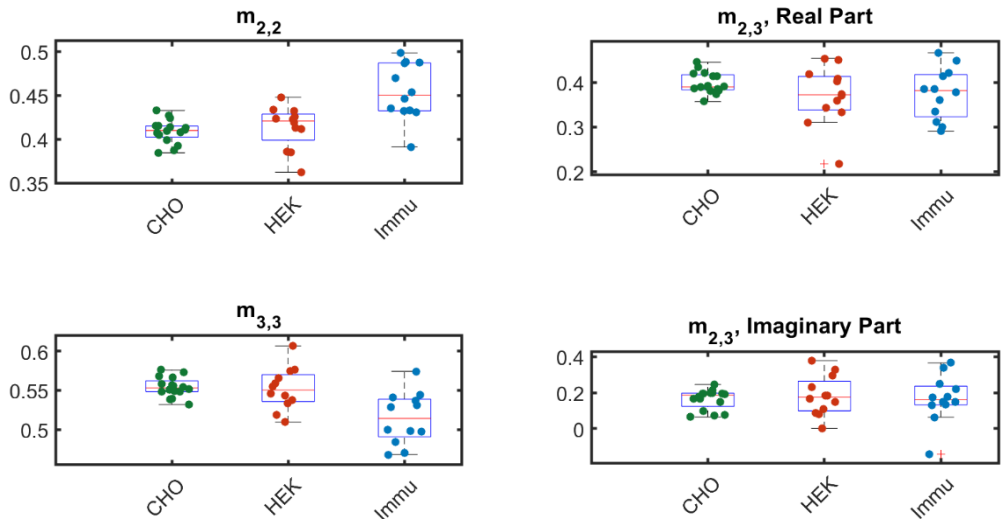

Figure 6. Box plot of elements $m_{2,2}$, $m_{2,3}$ (real and imaginary parts separately), and $m_{3,3}$ of all investigated types of samples: CHO, HEK, and Immu-Mount.

Figure 6 displays the repeatability of elements $m_{2,2}$, $m_{2,3}$-real, $m_{3,3}$, and $m_{2,3}$-imaginary parts separately of all investigated types of samples. In order to further differentiate cell cultures from the host medium, we utilized data of $m_{2,2}$ and $m_{3,3}$ for further Wilcoxon rank-sum test. Due to normalization, the sum of the elements $m_{2,2}$ and $m_{3,3}$ is 1, we picked the data of $m_{3,3}$ for all samples and received the p value of CHO with Immu-Mount and HEK with Immu-Mount, which can be seen in Table 5.

Table 5 Wilcoxon Rank-Sum Test P Values for CHO and HEK with Immu-Mount.

| Sample types | CHO | HEK |
|---|---|---|
| Immu-Mount | $2.24 \times 10^{-4}$ | 0.01 |

With the data of $m_{3,3}$ for all samples, the p value of HEK with Immu-Mount diminished to 0.01, which is considered to signify a notable difference between the two species. Here we successfully differentiate CHO, Immu-Mount and HEK one from all the others. This indicates that the measured polarization response is related rather to morphology of the cell in the studied monolayer cultures. With planned experiments on different density of cells in the samples and extended list of cell types, the presented results have a great potential for highly sensitive detection and differentiation of delicate samples like monolayer cell cultures.

# 4 CONCLUSION

Here we utilized polarization entangled photons pairs to assess the potential quantum polarimetry as a means to identify different cell mammalian cell types arranged in monolayers. We experimentally demonstrated the sensitivity of this method by distinguishing the response of a cell culture sample from the response of homogenous host medium. In addition, we could measure differences in the response of two types of monolayer cell cultures, Chinese hamster ovary and human embryonic kidney cells as well as their host material. This holds significant potential, as it is, to our knowledge, the first experiment that has the potential to enable monolayer cell culture differentiation using a polarimetry-based method. Our findings not only contribute to the advancement of biological sample detection techniques but also open new avenues for employing quantum-assisted methodologies in complex biological system analysis. Future experiments fully using the advantageous features of correlation-based measurement techniques based on photon pairs, such as enhanced SNR, can further extend the limits of this measurement approach.

**ACKNOWLEDGMENTS**

This work has been funded by the German Ministry of Education and Research (FKZ 13N14877, 13N16441), European Union's Horizon 2020 research and innovation programme (Grant Agreement No. 899580); and the Cluster of Excellence "Balance of the Microverse" (EXC 2051 – project 390713860). L. Zhang thanks for funding of this work also to China Scholarship Council for initiating the PhD exchange program, V. R. Besaga thanks for funding of this work also to ProChance-career program (AZ 2.11.3- A1/2022-01) of the Friedrich Schiller University Jena.

L. Zhang and V. R. Besaga have contributed equally to this work. L. Zhang and V. R. Besaga should be considered joint first author

AUTHOR CONTRIBUTIONS

L. Zhang conducted the experiments, collected and analyzed the data, wrote the manuscript, V. R. Besaga conceived the study, designed and implemented the experiments, analyzed the data, edited the manuscript. P. Rühl prepared the samples, analyzed the data, edited the manuscript. C. Zou analyzed the data and edited the manuscript. S. H. Heinemann, Y. Wang, and F. Setzpfandt supervised the research, edited the manuscript. All authors reviewed the manuscript.

CONFLICT OF INTEREST

The authors declare no financial or commercial conflict of interest.

DATA AVAILABILITY STATEMENT

The data that support the findings of this study are available from the corresponding author upon reasonable request.

REFERENCES

1. Gilaberte Basset, M., F. Setzpfandt, F. Steinlechner, E. Beckert, T. Pertsch, and M. Gräfe, *Perspectives for Applications of Quantum Imaging.* Laser & Photonics Reviews, 2019. **13**(10): p. 1900097.
2. Lee, S.-Y., Y.S. Ihn, and Z. Kim, *Quantum illumination via quantum-enhanced sensing.* Physical Review A, 2021. **103**(1).
3. Brambilla, E., L. Caspani, O. Jedrkiewicz, L.A. Lugiato, and A. Gatti, *High-sensitivity imaging with multi-mode twin beams.* Physical Review A, 2008. **77**(5).
4. Fox, A.M., *Quantum optics: an introduction.* Vol. 15. 2006: Oxford University Press, USA.
5. Kalashnikov, D.A., Z. Pan, A.I. Kuznetsov, and L.A. Krivitsky, *Quantum Spectroscopy of Plasmonic Nanostructures.* Physical Review X, 2014. **4**(1).
6. Li, T., F. Li, X. Liu, V.V. Yakovlev, and G.S. Agarwal, *Quantum-enhanced stimulated Brillouin scattering spectroscopy and imaging.* Optica, 2022. **9**(8): p. 959-964.
7. He, H., R. Liao, N. Zeng, P. Li, Z. Chen, X. Liu, and H. Ma, *Mueller Matrix Polarimetry—An Emerging New Tool for Characterizing the Microstructural Feature*

8. Tuniyazi, A., T. Mu, X. Jiang, F. Han, H. Li, Q. Li, H. Gong, W. Wang, and B. Qin, *Snapshot polarized light scattering spectroscopy using spectrally-modulated polarimetry for early gastric cancer detection.* J Biophotonics, 2021. **14**(9): p. e202100140.
9. Le Gratiet, A., A. Bendandi, C.J.R. Sheppard, and A. Diaspro, *Polarimetric optical scanning microscopy of zebrafish embryonic development using the coherency matrix.* J Biophotonics, 2021. **14**(6): p. e202000494.
10. Peyvasteh, M., A. Popov, A. Bykov, A. Pierangelo, T. Novikova, and I. Meglinski, *Evolution of raw meat polarization-based properties by means of Mueller matrix imaging.* J Biophotonics, 2021. **14**(5): p. e202000376.
11. Qi, J. and D.S. Elson, *Mueller polarimetric imaging for surgical and diagnostic applications: a review.* J Biophotonics, 2017. **10**(8): p. 950-982.
12. Horn, R. and T. Jennewein, *Auto-balancing and robust interferometer designs for polarization entangled photon sources.* Opt Express, 2019. **27**(12): p. 17369-17376.
13. James, D.F.V., P.G. Kwiat, W.J. Munro, and A.G. White, *Measurement of qubits.* Physical Review A, 2001. **64**(5).
14. Hauck, J.S., D. Moon, X. Jiang, M.-E. Wang, Y. Zhao, L. Xu, H. Quang, W. Butler, M. Chen, E. Macias, X. Gao, Y. He, and J. Huang, *Heat shock factor 1 directly regulates transsulfuration pathway to promote prostate cancer proliferation and survival.* Communications Biology, 2024. **7**(1).
15. Shalon, D., R.N. Culver, J.A. Grembi, J. Folz, P.V. Treit, H. Shi, F.A. Rosenberger, L. Dethlefsen, X. Meng, E. Yaffe, A. Aranda-Diaz, P.E. Geyer, J.B. Mueller-Reif, S. Spencer, A.D. Patterson, G. Triadafilopoulos, S.P. Holmes, M. Mann, O. Fiehn, D.A. Relman, and K.C. Huang, *Profiling the human intestinal environment under physiological conditions.* Nature, 2023. **617**(7961): p. 581-591.